\documentclass[aps,onecolumn,12pt,nofootinbib,showpacs]{revtex4}
\usepackage[dvips]{graphicx} \usepackage{amssymb} \usepackage{psfrag}

\begin{document}
\begin{flushright}
\vspace*{-1.5cm}
BA-06-18\\
\end{flushright}
\vspace{-.1cm}
\title{Supersymmetric And Smooth Hybrid Inflation \\ In The Light Of WMAP3}
\author{Mansoor ur Rehman} \email{rehman@udel.edu}
\author{V. N. {\c S}eno$\breve{\textrm{g}}$uz} \email{nefer@udel.edu}
\author{Qaisar Shaf\mbox{}i} \email{shafi@bxclu.bartol.udel.edu}
\affiliation{Bartol Research Institute, Department of Physics and Astronomy, University of Delaware, Newark, DE
19716, USA} 
\begin{abstract} 
In their minimal form both supersymmetric and smooth hybrid inflation
yield a scalar spectral index $n_s$ close to 0.98, to be contrasted with
the result $n_{s}=0.951_{-0.019}^{+0.015}$ from WMAP3. To realize better agreement,
following Ref. \cite{Bastero-Gil:2006cm}, we extend the parameter space of these models by employing
a non-minimal K\"ahler potential. We also discuss non-thermal leptogenesis by inflaton
decay and obtain new bounds in these models on the reheat temperature to explain the observed baryon
asymmetry.
\end{abstract}
\maketitle

\section{Introduction}
Supersymmetric (SUSY) hybrid inflation models
\cite{Dvali:1994ms,Lazarides:2001zd}, through their
connection to the grand unification scale, provide a compelling framework for
the understanding of the early universe. SUSY hybrid inflation is defined by 
the superpotential $W$ \cite{Copeland:1994vg,Dvali:1994ms}\footnote{This superpotential was considered
in the context of electroweak symmetry breaking  in Ref. \cite{Fayet:1974pd}.}
\begin{equation} \label{super1}
W=\kappa \widehat{S}(\widehat{\Phi }\widehat{\overline{\Phi }}-M^{2})\,,
\label{1}
\end{equation}
where $\widehat{S}$ is a gauge singlet and $\widehat{\Phi }$, $\ \widehat{%
\overline{\Phi }}$ are a conjugate pair of superfields transforming as
nontrivial representations of some group $G$. A simple example of the gauge group $G$ can be provided by the
standard model gauge group supplemented by a gauged $U(1)_{B-L}$, which
requires, from the anomaly cancellations, the presence of three right handed
neutrinos. The
K\"ahler potential can be expanded as
\begin{equation}
K=\left| S\right| ^{2}+\left| \Phi \right| ^{2}+\left| \overline{\Phi }%
\right| ^{2}+\kappa _{S}\frac{\left| S\right| ^{4}}{4m_{p}^{2}}+\cdots  \label{kah}
\end{equation}
where $S,$ $\Phi $ and $\overline{\Phi }$ are the bosonic components of the
superfields, and $m_{p}=2.4\times 10^{18}$ GeV is the reduced Planck mass.

In these models, if the K\"ahler potential is assumed to be minimal, the scalar
spectral index $n_s\approx0.985$ for the dimensionless coupling $\kappa$ in the superpotential
$\sim10^{-2}$, and larger for other values of $\kappa$. The running of the spectral index
${\rm d}n_s/{\rm d}\ln k$ and the tensor to scalar ratio $r$ is negligible
\cite{Senoguz:2003zw,Senoguz:2004vu,Jeannerot:2005mc}. On the other hand, for negligible $r$ 
the WMAP three year central value for the spectral index is 
$n_{s}\approx0.95$, and SUSY hybrid inflation with a minimal K\"ahler potential
is disfavoured at a $2\sigma$ level \cite{Spergel:2006hy}.\footnote{Note
however that it is claimed the error contours are in fact considerably larger
than shown in Ref. \cite{Spergel:2006hy}, with $n_s\approx0.985$ only
disfavoured at a $1\sigma$ level \cite{Kinney:2006qm}.}

It was recently shown that the spectral index for SUSY hybrid inflation can be
substantially modified in the presence of a small negative mass term in the
potential. This can result from a non-minimal K\"ahler potential, in particular
from the term proportional to the dimensionless coupling $\kappa_S$ above
\cite{Bastero-Gil:2006cm}. Ref. \cite{Bastero-Gil:2006cm} 
presents the results for $\kappa$ values
$\gtrsim10^{-3}$. In this paper we will explore the possible extension
of the range of $\kappa$ to lower values depending on $\kappa_{S}$. 
As we will see increasing the value of $\kappa _{S}$
increases the range of $\kappa$ to lower values, consistent with
the measured value of the curvature perturbation amplitude $\mathcal{R}=4.86\times 10^{-5}$.
This in turn extends
the range of other parameters like the symmetry breaking scale $M$,
the inflaton mass $m_{\rm inf}$ and the reheat temperature $T_{r}$.

The outline of the paper is as follows. 
In section \ref{susyhyb} we consider 
SUSY hybrid inflation with a non-minimal
K\"ahler potential. Using the standard constraints, we present our
numerical results for the allowed range of $\kappa$, $n_s$ and $M$
for different values of $\kappa_S$. In section \ref{smoothhyb} we
consider smooth hybrid inflation, an extension of SUSY hybrid inflation
which evades potential problems associated with topological defects. 
We again present how the parameters change with $\kappa_S$.
In section \ref{reheat} we discuss non-thermal leptogenesis by inflaton
decay and show that enough matter asymmetry can be generated with 
lower values of reheat temperature for nonzero $\kappa_S$ in both SUSY and smooth hybrid
inflation. We then conclude by reviewing our results in section \ref{conc}.

\section{SUSY hybrid inflation with non-minimal K\"ahler potential}\label{susyhyb}
Non-minimal supersymmetric hybrid inflation may be defined by the
superpotential given in Eq. (\ref{super1}), together with a general
K\"ahler potential
\begin{equation}
K=\left| S\right| ^{2}+\left| \Phi \right| ^{2}+\left| \overline{\Phi }%
\right| ^{2}+\kappa _{S}\frac{\left| S\right| ^{4}}{4m_{p}^{2}}+\kappa
_{S\phi }\frac{\left| S\right| ^{2}\left| \Phi \right| ^{2}}{m_{p}^{2}}%
+\kappa _{S\overline{\phi }}\frac{\left| S\right| ^{2}\left| \overline{\Phi }%
\right| ^{2}}{m_{p}^{2}}+\kappa _{SS}\frac{\left| S\right| ^{6}}{6m_{p}^{4}}%
+\cdots \label{2}
\end{equation}
The SUGRA scalar potential is given by
\begin{equation}
V_{F}=e^{K/m_{p}^{2}}\left(
K_{ij}^{-1}D_{z_{i}}WD_{z^{*}_j}W^{*}-3m_{p}^{-2}\left| W\right| ^{2}\right)
\label{3}
\end{equation}
with $z_{i}$ being the bosonic components of the superfields $\widehat{z}%
_{i}\in \{\widehat{\phi },\widehat{S},\cdots\}$ and where we have defined
\begin{eqnarray*}
D_{z_{i}}W &\equiv &\frac{\partial W}{\partial z_{i}}+m_{p}^{-2}\frac{%
\partial K}{\partial z_{i}}W \\
K_{ij} &\equiv &\frac{\partial ^{2}K}{\partial z_{i}\partial z_{j}^{*}}
\end{eqnarray*}
and $D_{z_{i}^{*}}W^{*}=\left( D_{z_{i}}W\right) ^{*}.$ 
In the D-flat direction $|\Phi|=|\overline{\Phi}|$, 
and using Eqs. (\ref{1}, \ref{2}) in Eq. (\ref{3}), we get 
\cite{Panagiotakopoulos:1997qd,Asaka:1999jb}
\[
V_{F}= \kappa ^{2}M^{4}\left( 1-\kappa _{S}\frac{\left| S\right| ^{2}}{%
m_{p}^{2}}+\gamma _{S}\frac{\left| S\right| ^{4}}{2m_{p}^{4}}+\cdot \cdot
\cdot \right) +\kappa ^{2}\left| \Phi \right| ^{2}\left( 2\left( \left|
S\right| ^{2}-M^{2}\right) +\cdots \right) +\cdots 
\]
where $\gamma _{S}=1-\frac{7\kappa _{S}}{2}+2\kappa _{S}^{2}-3\kappa _{SS}.$%

In the following discussion and calculations, we will set all the 
couplings in the K\"ahler potential except $\kappa_S$ to zero. The only
coupling except $\kappa_S$ which could have a significant effect is $\kappa_{SS}$, since
if $\kappa_{SS}$ is large and positive the quartic term becomes negative.
The potential in this case is lifted by a higher order term for large
values of $\kappa$.

Assuming suitable initial conditions the fields get trapped in the inflationary
valley of local minima at $\left| S\right| >S_{c}=M$ and $\left| \Phi
\right| =\left| \overline{\Phi }\right| =0$, where $G$ is unbroken. The
potential is dominated by the constant term $V_{0}=\kappa ^{2}M^{4}$. 
Inflation ends when the inflaton drops below its
critical value $S_{c}=M$ and the fields roll towards the global SUSY minimum of
the potential $\left| S\right| =0$ and $\left| \Phi \right| =\left| 
\overline{\Phi }\right| =M$. In the inflationary trajectory the potential is 
\[
V_{F}= \kappa ^{2}M^{4}\left( 1-\kappa _{S}\frac{\left| S\right| ^{2}}{%
m_{p}^{2}}+\gamma _{S}\frac{\left| S\right| ^{4}}{2m_{p}^{4}}+\cdots \right)\,.
\]
Taking also into account the radiative correction \cite{Dvali:1994ms} and soft SUSY breaking terms,
the potential is of the following form 
\begin{eqnarray}
V &\approx&V_{F}+\Delta V_{1loop}+V_{soft}  \label{4} \\
&= &\kappa ^{2}M^{4}\left( 1-\kappa _{S}\left( \frac{M}{m_{p}}\right)
^{2}x^{2}+\gamma _{S}\left( \frac{M}{m_{p}}\right) ^{4}\frac{x^{4}}{2}+\frac{%
\kappa ^{2}\mathcal{N}}{8\pi ^{2}}F\right) +a\kappa M^{3}x+a^{2}M^{2}x^{2}
\end{eqnarray}
where 
\[
\Delta V_{1loop}=\frac{1}{64\pi ^{2}}Str[{\cal M}^{4}(S )(\ln \frac{{\cal M}^{2}(S)%
}{Q^{2}}-\frac{3}{2})]=\frac{\left( \kappa M\right) ^{4}}{8\pi ^{2}}\mathcal{%
N}F(x) 
\]
and 
\[
V_{soft}=a\kappa M^{3}x+a^{2}M^{2}x^{2} 
\]
with
\begin{equation}
F(x)=\frac{1}{4}\left( \left( x^{4}+1\right) \ln \frac{\left( x^{4}-1\right) 
}{x^{4}}+2x^{2}\ln \frac{x^{2}+1}{x^{2}-1}+2\ln \frac{\kappa ^{2}M^{2}x^{2}}{%
Q^{2}}-3\right)
\end{equation}
and 
\begin{equation}
a=m_{3/2}2\left| 2-A\right| \cos [\arg S+\arg (2-A)].
\end{equation}
Here $\mathcal{N}$ is the dimensionality of the representation of the fields 
$\Phi$ and $\overline{\Phi },$ $Q$ the renormalization scale and $x=\left|
S\right| /M.$ In our numerical calculations we will take $a=1$ TeV.

The number of $e$-folds after the comoving scale $l$ has crossed the horizon
is given by
\begin{equation}
N_{l}=2\left( \frac{M}{m_{p}}\right) ^{2}\int_{1}^{x_{l}}\left( \frac{V}{%
\partial _{x}V}\right) dx  \label{5}
\end{equation}
where $|S_{l}|=x_{l}M$ is the value of the field at the comoving scale $l$.
During inflation, the comoving
scale corresponding to $k_{0}=0.002$ Mpc$^{-1}$ exits the horizon
at approximately
\begin{equation}
N_{0}=53+\frac{1}{3}\ln \left( \frac{T_{r}}{10^{9}~{\rm GeV}}\right) +\frac{2}{3}%
\ln \left( \frac{\sqrt{\kappa}M}{10^{15}~{\rm GeV}}\right)   \label{7}
\end{equation}
where $T_{r}$ is the reheating temperature, and the subscript `0' 
indicates that the values are taken at $k_{0}$.

The amplitude of the curvature perturbation is given by
\begin{equation}
\mathcal{R}=\frac{M}{\sqrt{6}\pi m_{p}^{3}}\left( \frac{V^{3/2}}{\partial
_{x}V}\right) _{x=x_{0}}=4.86\times 10^{-5}  \label{6}
\end{equation}
which is the WMAP normalization at $k_{0}$ \cite{Spergel:2006hy}.

\begin{figure}[htb] 
\includegraphics[angle=0, width=13cm]{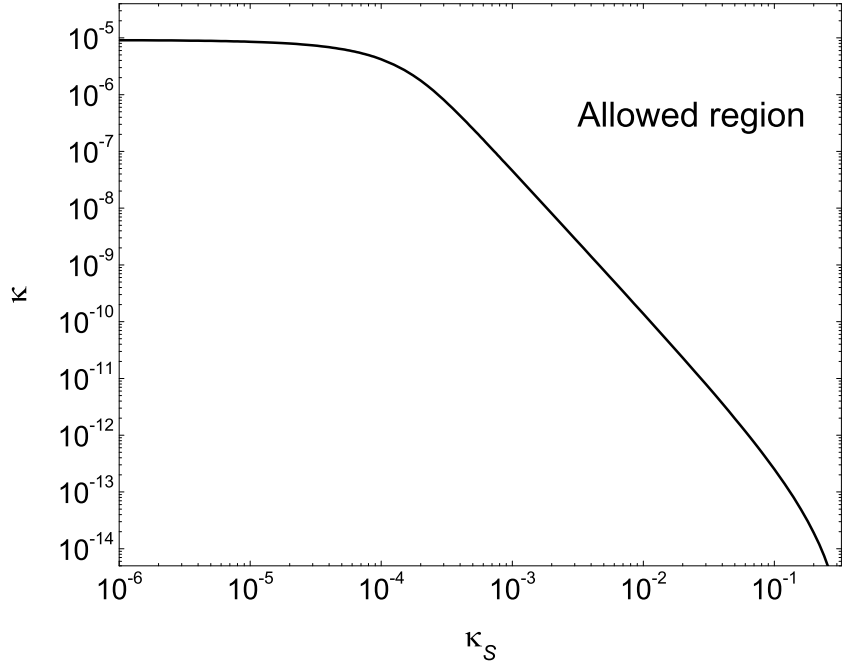}
\vspace{-1cm} 
\caption{The region in the $\kappa$ and $\kappa_S$ plane
satisfying $\mathcal{R}=4.86\times10^{-5}$.} \label{lower}
\end{figure}

For small values of $\kappa ,$ $x_{0}$ becomes practically equal to $1$ 
and the radiative term becomes negligible. The soft mass term $a^2 M^2 x^2$
is likewise negligible. Eq. (\ref{6}) then yields 
\begin{equation} \label{mathr}
\mathcal{R}=\frac{\kappa ^{2}}{\sqrt{6}\pi }\left( \frac{M}{m_{p}}\right)
^{4}\left( \frac{1}{2\kappa \gamma _{S}\left( \frac{M}{m_{p}}\right)
^{5}-2\kappa \kappa _{S}\left( \frac{M}{m_{p}}\right) ^{3}+\frac{a}{m_{p}}}%
\right) .
\end{equation}
Maximizing this expression with respect to $M$ gives us the
lower bound on $\kappa $ from $\mathcal{R}_{\max }=4.86\times 10^{-5}.$
For small values of $\kappa _{S}$ the quartic term is dominant over the quadratic
term and these terms become equal for $\kappa _{S}\sim 2\times 10^{-4}.$ For
greater values of $\kappa _{S}$ the quadratic term becomes
dominant. Numerically, we obtain the lower bounds on $\kappa $ as shown in Fig. \ref{lower}, with
\begin{eqnarray*}
\kappa  &\gtrsim &c_{1}^{5/6}\left( 1-c_{1}^{1/3}c_{2}\right) ^{5/6}\text{%
for }\kappa _{S}<6\times 10^{-5}\,, \\
\kappa  &\gtrsim &\left( \frac{1}{c_{2}}\right) ^{5/2}\left( 1-\frac{1}{%
c_{2}^{3}c_{1}}\right) ^{5/2}\text{for }\kappa _{S}>6\times 10^{-5}\,, \\
\kappa  &\gtrsim &9.1\times 10^{-6}\text{ for }\kappa _{S}\sim 6\times
10^{-5}\,, \\
\text{where }c_{1} &=&\frac{5b\sqrt{6}\pi \mathcal{R}_{\max }}{\left( \frac{%
2b}{\gamma _{S}}\right) ^{4/5}}\text{ , }c_{2}=\frac{4\kappa _{S}}{5b}\left( 
\frac{2b}{\gamma _{S}}\right) ^{3/5}\text{ and }b=\frac{a}{m_{p}}\,.
\end{eqnarray*}
\clearpage

\begin{figure}[t] 
\includegraphics[angle=0, width=13cm]{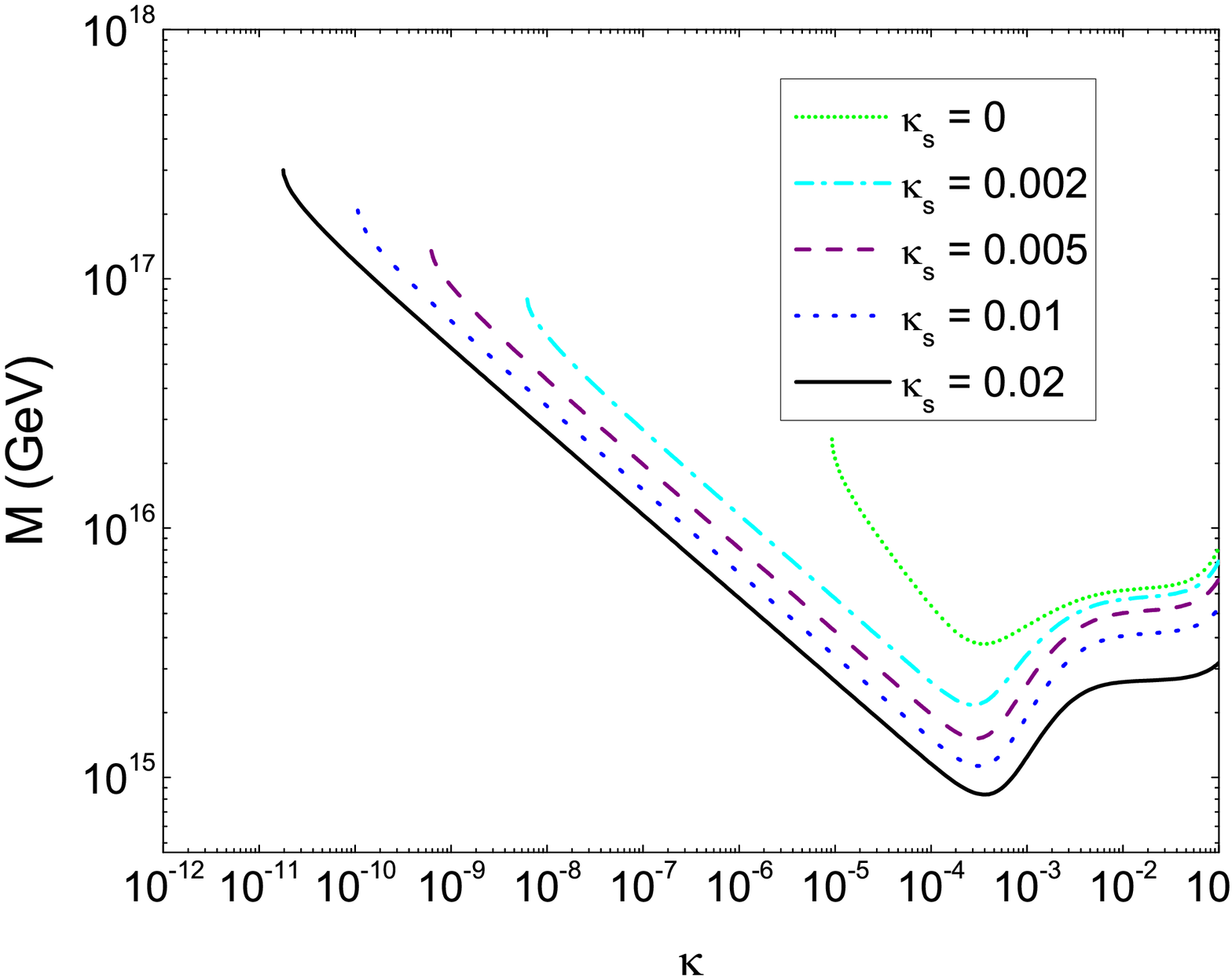} 
\vspace{-1cm}
\caption{$M$ as a function of $\kappa $ for different values of $\kappa
_{S}$ ($\mathcal{N}=1$).} \label{mvsk}
\end{figure}

For values of $\kappa $ and $M$ such that we can ignore both the quartic
and the loop terms in the potential, $\mathcal{R}$ becomes\footnote{%
Here we again have $x_{0}\approx 1$.} 
\begin{equation} \label{mathr2}
\mathcal{R}\approx \frac{\kappa ^{2}}{\sqrt{6}\pi }\left( \frac{M}{m_{p}}%
\right) ^{4}\left( \frac{1}{-2\kappa \kappa _{S}\left( \frac{M}{m_{p}}%
\right) ^{3}+\frac{a}{m_{p}}}\right)\,,
\end{equation}
which for $\kappa _{S}>0$ gives the expression for $M$:
\[
M\approx \left( \frac{b}{2\kappa \kappa _{S}}\right) ^{1/3}\left( 1-\frac{%
\kappa ^{2}}{b\sqrt{6}\pi \mathcal{R}}\left( \frac{b}{2\kappa \kappa _{S}}%
\right) ^{4/3}\right) ^{1/3}m_{p}\,.
\]
Maximizing Eq. (\ref{mathr}) with respect to $M$, we find $M=(2b/\kappa\gamma_S)^{1/5}m_p$
at the lower bound on $\kappa$.
The numerical values of $M$ obtained using Eqs. (\ref{4}--\ref{7})
is shown in Fig. \ref{mvsk}.

\begin{figure}[t] 
\includegraphics[angle=0, width=13cm]{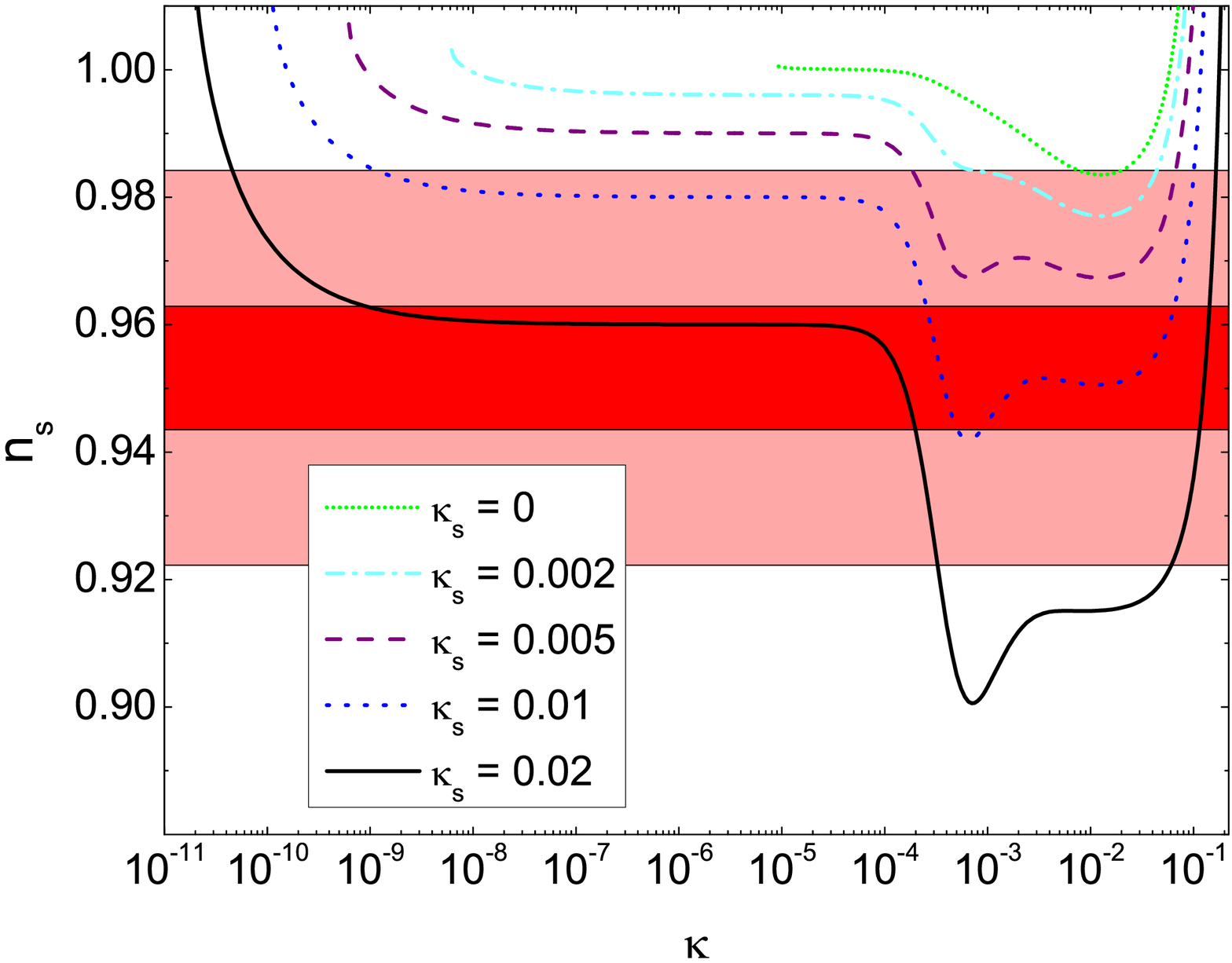} 
\vspace{-1cm} 
\caption{$n_{s}$ as a function of $\kappa $ for different values of $\kappa _{S}$ ($%
\mathcal{N}=1$). The red and pink bands correspond to the WMAP $1\sigma$ and
2$\sigma$ range \cite{Spergel:2006hy}.} \label{nvsk}
\end{figure}

\begin{figure}[t] 
\includegraphics[angle=0, width=12.8cm]{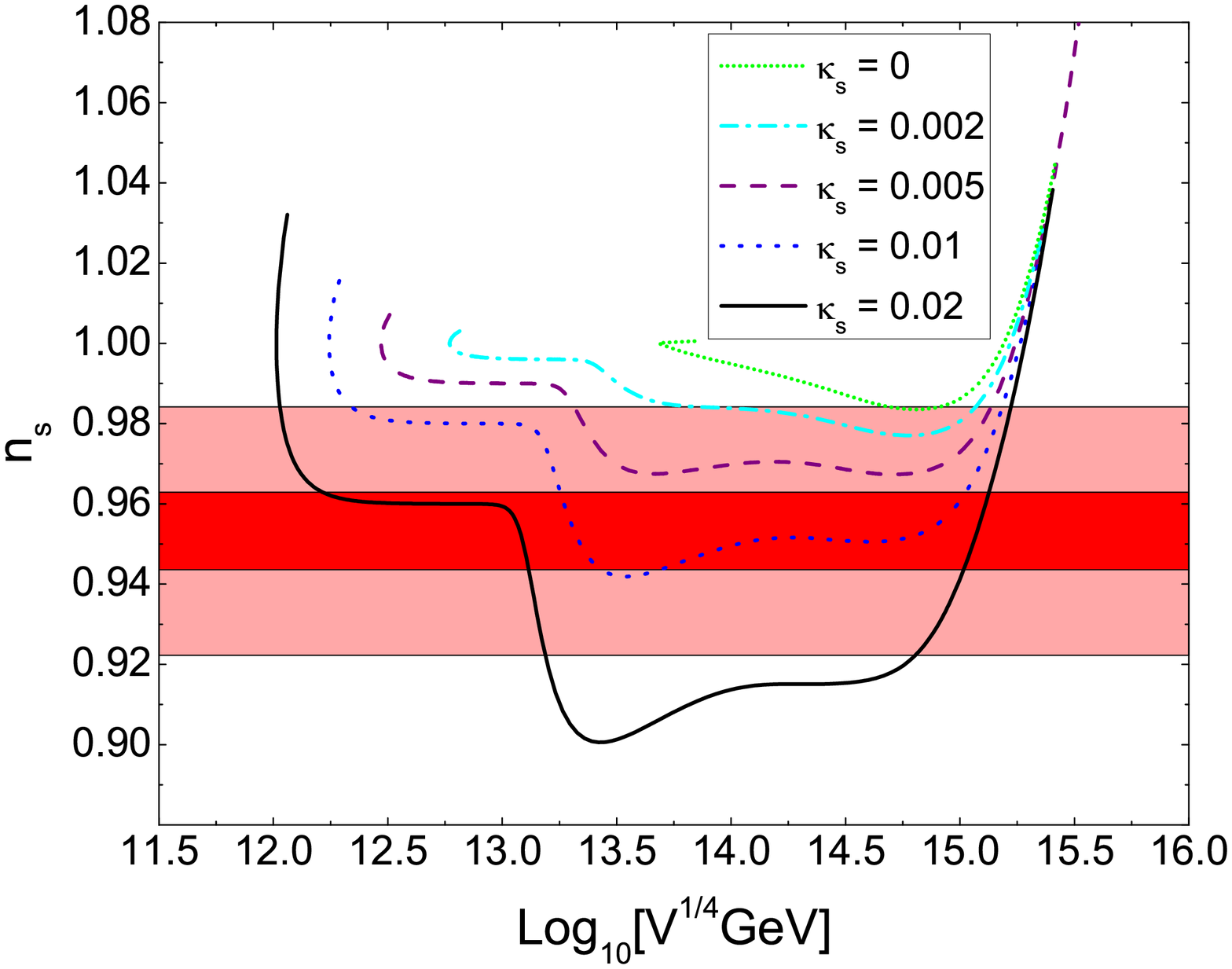} 
\vspace{-1.1cm} 
\caption{$n_{s}$ as a function of $\log[V^{1/4}{\rm GeV}]$ for different values of $\kappa _{S}$ ($%
\mathcal{N}=1$). The red and pink bands correspond to the WMAP $1\sigma$ and
2$\sigma$ range \cite{Spergel:2006hy}.} \label{nsvlog}
\end{figure}

\begin{figure}[b] 
\includegraphics[angle=0, width=12.8cm]{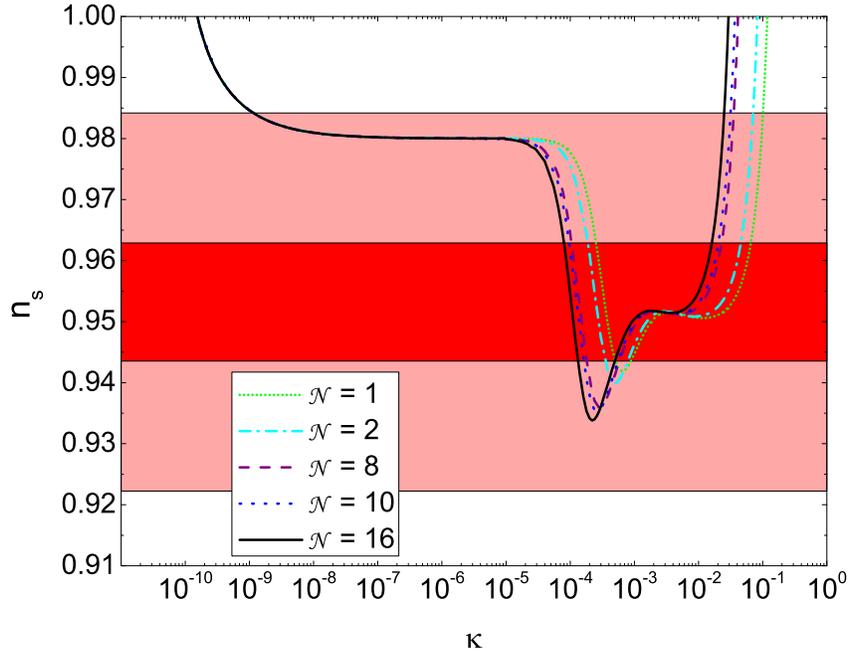} 
\vspace{-1.1cm} 
\caption{$%
n_{s}$ as a function of $\kappa $ for different values of $\mathcal{N}$ ($%
\kappa _{S}=0.01$). The red and pink bands correspond to the WMAP $1\sigma$ and
2$\sigma$ range \cite{Spergel:2006hy}.} \label{nsvkforn}
\end{figure}

The slow-roll parameters may be defined as
\begin{equation}
\epsilon =\frac{m_{p}^{2}}{2}\left( \frac{V^{\prime} }{V}\right) ^{2}\,, \quad
\eta =m_{p}^{2}\left( \frac{V^{\prime \prime} }{V}\right)  \nonumber\,, \quad
\xi ^{2} =m_{p}^{4}\left( \frac{V^{\prime} V^{\prime \prime\prime} }{V^{2}}\right)\,,
\end{equation}
where $V^{\prime}$ denotes the derivative with respect to the normalized real field
$\sigma\equiv\sqrt{2}Mx$.
Assuming the
slow-roll approximation is valid (i.e. $\epsilon\ll1,\eta\ll1$), the spectral index $%
n_{s}$ and the running of the spectral index ${\rm d}n_s/{\rm d}\ln k$ are
given by
\begin{eqnarray}
n_{s} &\approx &1-6\epsilon +2\eta  \label{8} \,,\\
\frac{dn_{s}}{d\ln k} &\approx &16\epsilon \eta -24\epsilon ^{2}-2\xi
^{2}\,.
\end{eqnarray}
Using Eq. (\ref{8}), we calculate $n_{s}$ as a function of $\kappa $ for different values of
 $\kappa _{S}$ (Fig. \ref{nvsk}).
In the $\kappa$ range where the quartic and loop terms
are subdominant, from Eq. (\ref{mathr2}) $n_s$ is approximated
to be 
\[
n_{S}=1-2\kappa _{S}+6\gamma _{S}\left( \frac{M}{m_{p}}\right)
^{2}x_{0}^{2}+\left( \frac{m_{p}}{M}\right) ^{2}\left( \frac{\kappa ^{2}%
\mathcal{N}}{8\pi ^{2}}\right) \left. \partial _{x}^{2}F\right|
_{x=x_{0}}\simeq 1-2\kappa _{S}\,.
\]
This range is represented by the horizontal sections in Fig. \ref{nvsk}. For still smaller
values of $\kappa$, the quartic term becomes important and the curves bend upward,
with $n_{s}$ becoming greater than $1$.\footnote{There is also an upper branch of
solutions for $M$ and $n_s$ as functions of $\kappa$,
where $n_s$ remains $>1$ \cite{Senoguz:2004vu}. We do not display this branch of solutions since
it is disfavoured by the WMAP results.} We also plot $n_s$ versus $\log[V^{1/4}{\rm GeV}]$ 
and $n_s$ for different values of ${\cal N}$ in Figs.
\ref{nsvlog}, \ref{nsvkforn}. The running of the spectral index
is negligible in SUSY hybrid inflation, with $|{\rm d}n_s/{\rm d}\ln k|\lesssim10^{-3}$.

\clearpage

\enlargethispage{\baselineskip}
\clearpage
\begin{figure}[t] 
\includegraphics[angle=0, width=13cm]{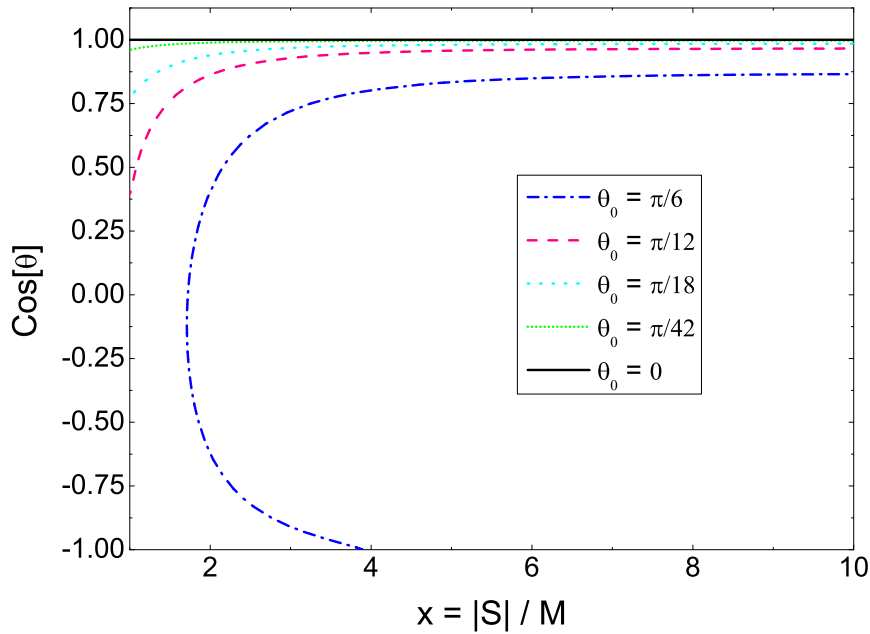} 
\vspace{-1cm}
\caption{An example to show that the change in $\theta\equiv$ $%
\arg S$ can be controlled by taking suitable initial conditions, $\theta
=\theta _{0}$ and $x=10x_{0}$ ($\arg (2-A)=0$, $\kappa =10^{-4}$,
$\kappa _{S}=0$).} \label{argS}
\end{figure}
It should be noted that for  large enough values of $\kappa_S$ or small enough
values of $\kappa$, the potential develops a false minimum at
$\sigma>\sigma_0$. Successful inflation then requires the $\sigma$ field to
have just enough kinetic energy so that it reaches the local maximum of the
potential with negligible kinetic energy. This seems rather improbable since it
is realized only for a very narrow band of initial
values.\footnote{Alternatively, if the field is trapped in the false minimum it
should tunnel to a point just beyond the local maximum, but this is exceedingly
improbable.} Furthermore, we assume that the $A$-term ($a\kappa M^{3}x$) in the
potential is positive. Since this depends on the value of $\theta\equiv{\rm arg}S$,
it should be checked whether the change in $\theta$ is small. As displayed in Fig.
\ref{argS}, this is possible but again requires specific initial
conditions.\footnote{For large values of $\kappa$
($\kappa\gtrsim5\times10^{-4}$ for $a=1$ TeV), the $A$-term in the potential does
not play a significant role. However, for smaller values of $\kappa$, the $A$-term
is important and its derivative determines $\mathcal{R}$ together with the derivative of the radiative term.
If the $A$-term is negative it should be subdominant with respect to the radiative
term. Since the derivative of the radiative term has a lower bound at $x=1$ depending
on $\kappa$, this condition puts a lower bound on $\kappa\approx3\times10^{-4}$ ($5\times10^{-4}$)
for $\kappa_S=0$ ($\kappa_S=0.01$).
There is also a different branch of solutions
with higher values of $M$
where instead of the radiative term the quartic term is important. These solutions
allow smaller values of $\kappa$, but with the quartic term dominating
the spectral index is greater than unity.}

Here some discussion of initial conditions is in order.
The initial values of the fields can vary in different regions of the universe.
Furthermore, the couplings in the K\"ahler potential are determined by the vacuum expectation
values (VEVs) of moduli fields, which
can also vary in different regions. The regions with VEVs such that
the inflaton mass is suppressed will inflate more and become exponentially large
compared to other regions. In this sense, for negative values of $\kappa_S$ so that
the potential has a positive (mass)$^2$ term, the smallness of $|\kappa_S|$ ($\kappa_S\ll1$)
can be regarded as a selection effect (see Ref. \cite{Ross:1995dq} for a discussion).\footnote{It
is worth noting that inflation can be realized using only the MSSM fields, with an apparent
tuning of parameters that can be similarly justified \cite{Allahverdi:2006iq}.} 

However, for positive values of $\kappa_S$ the potential has a negative (mass)$^2$ term
which can lead to a local maximum. Once the inflaton field is sufficiently close to this local maximum (with
negligible kinetic energy), eternal inflation is realized. It would then seem that
the regions satisfying the conditions for eternal inflation would always dominate,
since even if they are initially rare, their volume
will increase indefinitely \cite{Vilenkin:1983xq}.
It is, however, also possible that there are no regions satisfying these conditions.
Alternatively, eternal inflation could occur not only close to the local maximum
mentioned above, but also at higher energies regardless of the value of $\kappa_S$.
It then becomes notoriously difficult, if not impossible, to compute the probability distribution of
observables such as $n_s$, even if the initial distribution of $\kappa_S$ is
assumed to be known.\footnote{See Ref. \cite{Vilenkin:2006xv} for a recent review
of progress in defining probabilities in an eternally inflating spacetime, and 
Ref. \cite{Tegmark:2004qd} for discussion and computational examples.}

To summarize, it is not clear
whether the parameter range explored in this paper is less likely to be
observed compared to the minimal K\"ahler potential or negative $\kappa_S$ cases.
Even if we only consider $\kappa_S$ small enough so that the potential remains monotonic,
$n_s$ can still be significantly lower compared to the minimal
K\"ahler case for large values of $\kappa$, with $n_s\simeq0.95$ for $\kappa\gtrsim0.1$
and $\kappa_S\simeq\kappa/9$.

Finally, we note that for SUSY hybrid inflation there are additional constraints
if the symmetry breaking pattern produces cosmic strings \cite{Jeannerot:2005mc}.
For example, strings are produced when $\Phi,\,\overline{\Phi}$ break $U(1)_R\times U(1)_{B-L}$ to $U(1)_Y\times Z_2$ 
matter parity, but not when $\Phi,\,\overline{\Phi}$ are $SU(2)_R\times U(1)_{B-L}$ doublets.
In this section we assumed that cosmic strings are not produced.

\section{Smooth hybrid inflation} \label{smoothhyb}
A variation on SUSY hybrid inflation is obtained by imposing a
$Z_{2}$ symmetry on the superpotential, so that only even powers of the
combination $\Phi\overline{\Phi}$ are allowed \cite{Lazarides:1995vr,Senoguz:2004vu}:
\begin{equation} \label{super3} W=S\left(-v^2
+\frac{(\Phi\overline{\Phi})^{2}}{M^{2}_{*}}\right)\,, \end{equation}
\noindent where the dimensionless parameter $\kappa$ is absorbed in $v$.  The resulting scalar
potential possesses two (symmetric) valleys of local minima which are suitable
for inflation and along which the GUT symmetry is broken. The
inclination of these valleys is already non-zero at the classical level and the
end of inflation is smooth, in contrast to SUSY hybrid inflation. An important consequence is that
potential problems associated
with topological defects are avoided. This `smooth hybrid inflation'
model is similar to the `mutated hybrid inflation' model considered in Ref. \cite{Stewart:1995pt}
and generalized in Ref. \cite{Lyth:1996kt}.

The common VEV at the SUSY minimum
$M=\big|\langle\nu^c_H\rangle\big|=\big|\langle\overline{\nu}^c_H\rangle\big|=
(v\,M_*)^{1/2}$. For $\sigma^{2}\gg M^2$, the inflationary potential
is given by
\begin{equation} \label{v3}
V\approx v^{4}\left[1-\frac{2}{27}\frac{M^4}{\sigma^{4}}+\frac{\sigma^4}{8m^4_p}\right]\,,
\end{equation}
\noindent where the last term arises from the SUGRA correction for a minimal
K\"ahler potential \cite{Senoguz:2004vu}. The
soft terms in this case do not have a significant effect on the inflationary dynamics. If we
set $M$ equal to the SUSY GUT scale $M_{\rm GUT}=2\times10^{16}$ GeV, we get $v\approx1.4\times10^{15}$
GeV and $M_*\approx2.8\times10^{17}$ GeV. (Note that, if we express
Eq. (\ref{super3}) in terms of the coupling parameter $\kappa$,
this value corresponds to $\kappa\sim O(v^2/M^2_{\rm GUT})\sim 10^{-2}$.) 
The value of the field $\sigma$ is $1.1\times10^{17}$ GeV at
the end of inflation (corresponding to $\eta=-1$) and
$\sigma_0\approx2.4\times10^{17}$ GeV at $k_0$.  In the absence of the SUGRA
correction (which is small for $M\lesssim10^{16}$ GeV), $\sigma_0\propto
M^{2/3}\,m_p^{1/3}$, $\mathcal{R}\propto M^{10/3}/(M^2_*\,m_p^{4/3})$ and
the spectral index is given by \cite{Lazarides:1995vr}
\begin{equation} n_{s}\approx1-\frac{5}{3N_{0}}\approx0.97\,. \end{equation}
The SUGRA correction raises $n_{s}$
from 0.97 to above unity for
$M\gtrsim1.5\times10^{16}$ GeV \cite{Senoguz:2004vu}.

One problem with this model is that the cutoff scale $M_*$ is close to the
inflaton field value $\sigma_0$ for $M\simeq M_{\rm GUT}$. $M_*$ becomes
smaller than $\sigma_0$ for $M\lesssim10^{16}$ GeV, for which the effective
field theory is in general no longer valid. However, with a negative mass term that
could result from a non-minimal K\"ahler potential larger values of $M_*$ are possible.
Also, as in SUSY hybrid inflation, the spectral
index can have lower values.

\begin{figure}[t]
 \includegraphics[angle=0,
width=12cm]{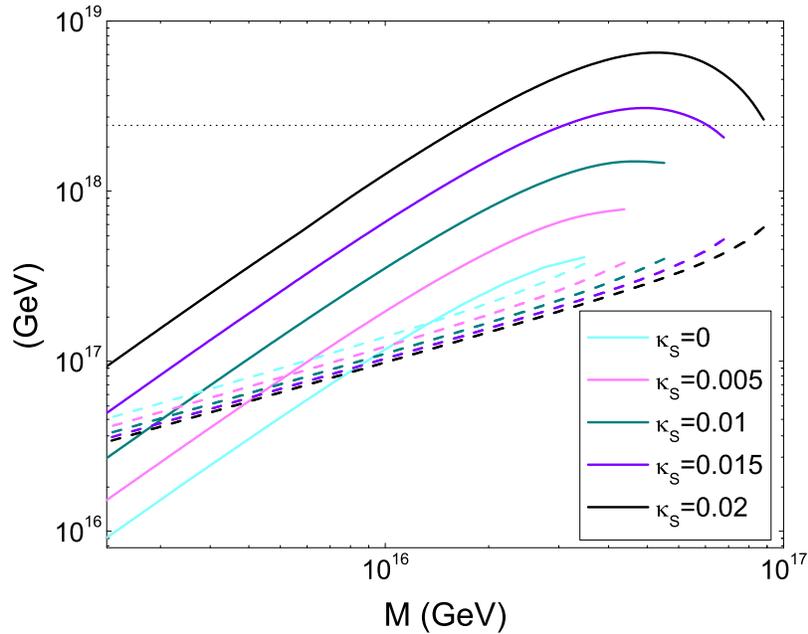} \vspace{-1cm} \caption{$M_*$ (solid) and $\sigma_0$
(dashed) as functions of the gauge symmetry breaking
scale $M$ for smooth hybrid inflation.} \label{nms1} 
\end{figure}

For a K\"ahler potential $K=|S|^2+|\Phi|^2+|\overline{\Phi}|^2+\lambda|S|^4/4M^2_*+\ldots$,
the potential is obtained as
\begin{equation} 
V\approx v^{4}\left[1-\frac{2}{27}\frac{M^4}{\sigma^{4}}-\frac{\kappa_S}{2}\frac{\sigma^2}{m^2_p}
+\frac{\gamma_S}{8}\frac{\sigma^4}{m^4_p}\right]\,.
\end{equation}
Here we have defined $\kappa_S\equiv \lambda m_p^2/M^2_*$ to express the potential
in a form similar to that of the previous section.
The $M_*$ and $\sigma_0$ values for different
values of $\kappa_S$ is displayed in Fig. \ref{nms1}.

\begin{figure}[t] \includegraphics[angle=0,
width=12cm]{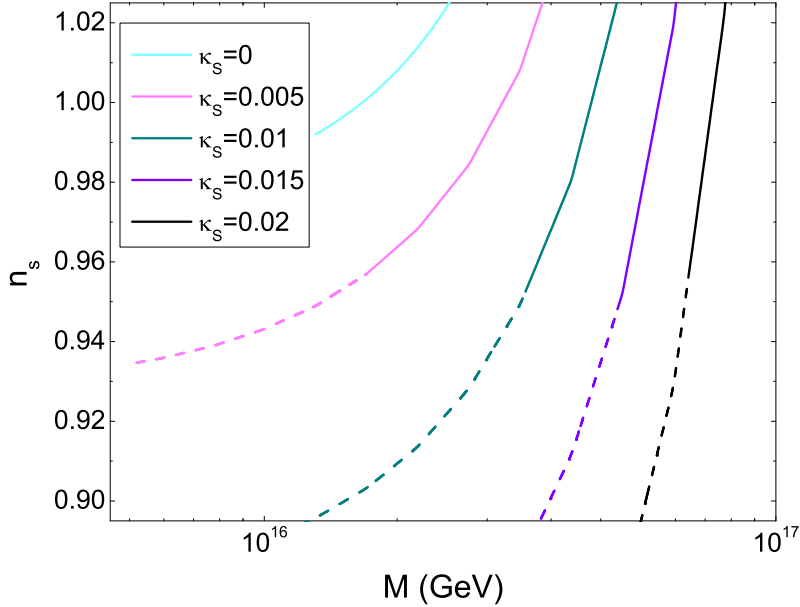} \vspace{-1cm} \caption[The spectral index $n_s$ 
as a function of the gauge symmetry breaking scale $M$ for smooth hybrid inflation.]{The spectral index $n_s$
as a function of the gauge symmetry breaking scale $M$ for smooth hybrid inflation. 
The dashed sections indicate that the field is
initially close to a local maximum.} \label{nms2} 
\end{figure}

The spectral index $n_s$ for different values of $\kappa_S$ is displayed in Fig. \ref{nms2}.
Note that for $\kappa_S=0$, requiring $\sigma_0<M_*$ constrains $n_s\gtrsim0.99$.
Having a non-zero $\kappa_S$ allows smaller values of $n_s$ in better
agreement with the WMAP3 results.
For large enough values of $\kappa_S$ or small enough values of $M$
(the dashed sections in the figure),
the potential develops a false minimum at $\sigma>\sigma_0$ as in SUSY hybrid inflation.
Again, even with $|\kappa_S|$ small enough so that there is no such false minimum,
$n_s$ can be as low as $0.95$.

\clearpage

\section{Reheat temperature and the gravitino constraint}\label{reheat}
After the end of inflation, the fields fall toward the SUSY vacuum and
perform damped oscillations about it. The VEVs of $\Phi $ and $\overline{%
\Phi }$, along their right handed neutrino components $\nu _{H}^{c}$, $%
\overline{\nu }_{H}^{c}$, break the gauge symmetry. The oscillating
system, which we collectively denote as $\chi$, consists of the two complex
scalar fields $\left( \delta \nu _{H}^{c}+\delta \overline{\nu }%
_{H}^{c}\right) /\sqrt{2}$ (where $\delta \nu _{H}^{c}$, $\delta \overline{%
\nu }_{H}^{c}$ are the deviations of $\nu _{H}^{c}$, $\overline{\nu }%
_{H}^{c} $ from $M$) and $S$, with equal mass $m_{\rm inf}$.%

We assume here that the inflaton $\chi$ decays predominantly into right
handed neutrino superfields $N_{i}$, via the superpotential coupling $%
(1/m_{P})\gamma_{ij}\overline{\Phi }\Phi N_{i}N_{j}$ or $\gamma_{ij}\overline{%
\Phi }N_{i}N_{j}$, where $i,j$ are family indices. Their subsequent out of
equilibrium decay to lepton and Higgs superfields generates lepton
asymmetry, which is then partially converted into the observed baryon
asymmetry by sphaleron effects.\footnote{Baryogenesis via leptogenesis was 
considered in Ref. \cite{Fukugita:1986hr}. Non-thermal leptogenesis
by inflaton decay was considered in Ref. \cite{Lazarides:1991wu},
and for SUSY hybrid inflation in Ref. \cite{Lazarides:1997dv}.}

The right handed neutrinos, as shown below, can be heavy compared to
the reheat temperature $T_r$.
Note that unlike thermal leptogenesis, there is then no washout factor 
since lepton number violating 2-body scatterings
mediated by right handed neutrinos are out of equilibrium as long as the
lightest right handed neutrino mass $M_1\gg T_r$ \cite{Fukugita:1990gb}.  More
precisely, the washout factor is proportional to $e^{-z}$ where $z=M_1/T_r$
\cite{Buchmuller:2003gz}, and can be neglected for $z\gtrsim10$. 
Without this assumption, generating sufficient lepton asymmetry
would require $T_r\gtrsim2\times10^9$ GeV
\cite{Giudice:2003jh}, and as discussed below
this is hard to reconcile with the gravitino constraint.

GUTs typically relate the Dirac neutrino masses to that of
the quarks or charged leptons. It is therefore reasonable to assume that the
Dirac masses are hierarchical. The low-energy neutrino data indicates that
the right handed neutrinos in this case will also be hierarchical in
general. As discussed in Ref. \cite{Akhmedov:2003dg}, setting the Dirac masses strictly equal
to the up-type quark masses and fitting to the neutrino oscillation
parameters generally yields strongly hierarchical right handed neutrino
masses ($M_{1}\ll M_{2}\ll M_{3}$), with $M_{1}$ $\sim 10^{5}$ GeV. The
lepton asymmetry in this case is too small by several orders of magnitude.
However, it is plausible that there are large radiative corrections to the
first two family Dirac masses, so that $M_{1}$ remains heavy compared to $%
T_{r}$. 

A reasonable mass pattern is therefore $M_{1}<M_{2}\ll M_{3}$,
which can result from either the dimensionless couplings $\gamma _{ij}$ or
additional symmetries. The dominant contribution to the lepton asymmetry is
from the decays with $N_{3}$ in the loop, as long as the first two
family right handed neutrinos are not quasi-degenerate. Under these
assumptions, the lepton asymmetry is given by \cite{Asaka:1999jb,Senoguz:2004vu}
\begin{equation}
n_{L}/s\lesssim 3\times 10^{-10}\frac{T_{r}}{m_{\rm inf}}\left( \frac{M_{i}}{%
10^{6}{\rm~GeV}}\right) \left( \frac{m_{\nu 3}}{0.05{\rm~eV}}\right) ,  \label{10}
\end{equation}
where $M_{i}$ denotes the mass of the heaviest right handed neutrino the
inflaton can decay into. 

\begin{figure}[t] 
\includegraphics[angle=0, width=13cm]{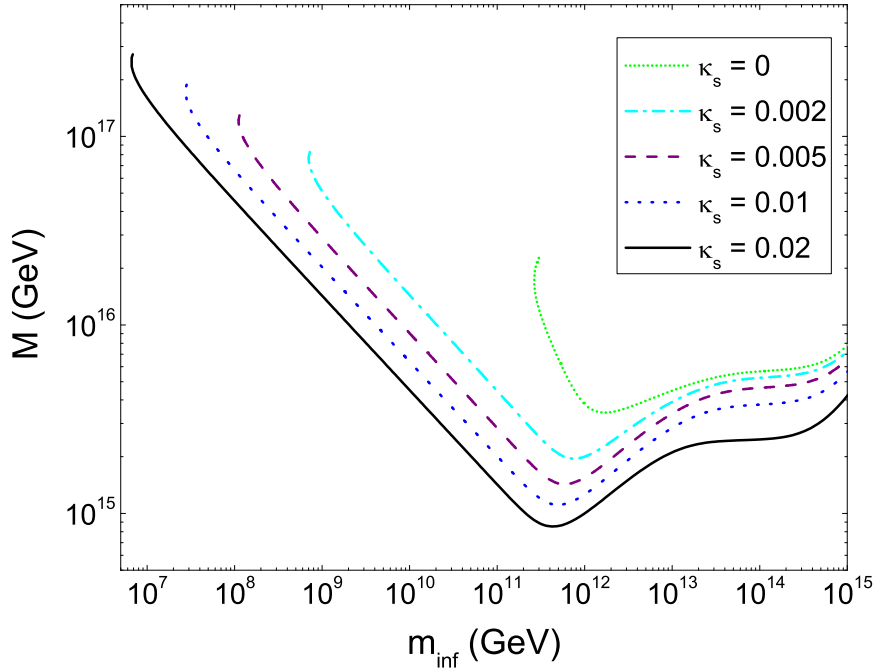} 
\vspace{-1cm} 
\caption{$M$ as a function of $m_{\rm inf}$
 for different values of $\kappa _{S}$ ($\mathcal{N}=1$).} \label{mvsinf}
\end{figure}

From the experimental value of the baryon to photon ratio $\eta _{B}\approx $ 
$6.1\times 10^{-10}$ \cite{Spergel:2006hy}, the required lepton asymmetry is found to be $%
n_{L}/s\approx 2.5\times10^{-10}$ \cite{Khlebnikov:1988sr}.
Since $m_{\rm inf}>2M_i$, Eq. (\ref{10}) then yields
\begin{equation} \label{trlimit} T_r\gtrsim1.6\times10^6\ {\rm GeV}\left(\frac{0.05\ {\rm
eV}}{m_{\nu3}}\right)\,. \end{equation}
This is a general bound valid for non-thermal leptogenesis by inflaton decay,
assuming hierarchical right handed neutrinos that are heavy compared to $T_r$.\footnote{Having 
quasi-degenerate neutrinos increases the lepton asymmetry per neutrino
decay $\epsilon$ \cite{Flanz:1996fb} and thus allows lower values of $T_r$ corresponding to lighter right
handed neutrinos. Provided that the neutrino mass splittings are comparable to
their decay widths, $\epsilon$ can be as large as $1/2$ \cite{Pilaftsis:1997jf}. 
The lepton asymmetry in this case is of order $T_r/m_{\rm inf}$ and sufficient lepton asymmetry 
can be generated with $T_r$ close to the electroweak scale.}
More specific bounds can be obtained using
the inflaton decay rate $\Gamma _{\chi }$ $=(1/8\pi
)(M_{i}^{2}/M^{2})m_{\rm inf}$. The reheat temperature $T_{r}$ is given by%
\begin{equation} \label{reheq}
T_{r}=\left( \frac{45}{2\pi ^{2}g_{*}}\right) ^{1/4}\left( \Gamma _{\chi
}m_{p}\right) ^{1/2}\approx 0.063\frac{(m_{p}m_{\rm inf})^{1/2}}{M}M_{i}\,.
\end{equation}
For SUSY hybrid inflation the values of $m_{\rm inf}=\sqrt{2}\kappa M$ are
shown in Fig. \ref{mvsinf}. Eq. (\ref{reheq}) yields the 
result that $M_i$ is about 200 (6) times heavier than
$T_r$, for $\kappa=10^{-5}$ ($10^{-2}$) with $\kappa_S=0$. $M_i/T_r$ decreases slightly
for non-zero $\kappa_S$, with $M_i/T_r\approx150$ (5) for the same $\kappa$ values and
$\kappa_S=0.01$. Thus, small values of $\kappa$ are consistent with ignoring washout
effects as long as the lightest right handed neutrino mass $M_1$ is also $\gg T_r$.

\begin{figure}[t] 
\includegraphics[angle=0, width=13cm]{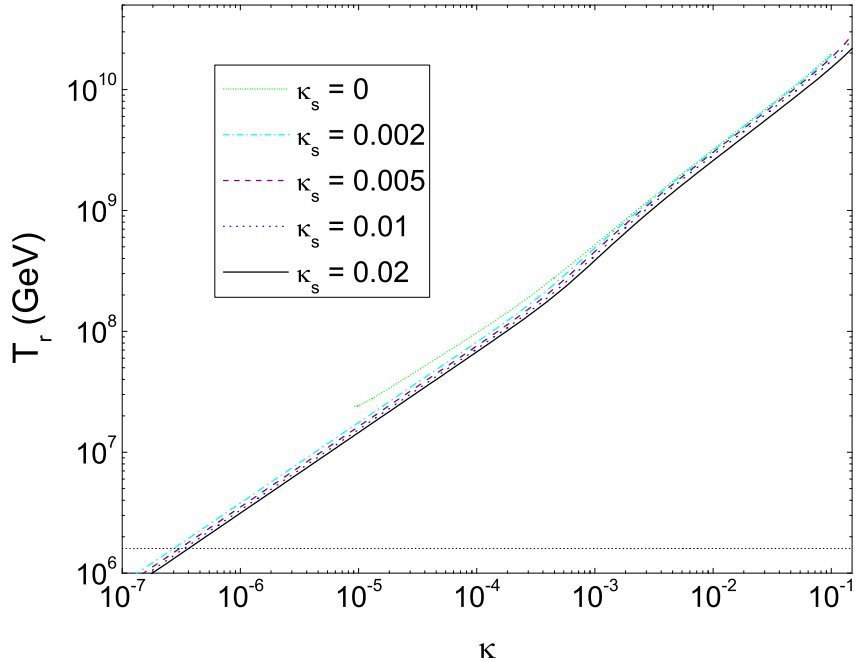}
 \vspace{-1cm} 
\caption{$T_{r}$ as a function of $\kappa $
for different values of $\kappa _{S}$ ($\mathcal{N}=1$).}
 \label{tvsk}
\end{figure}

Using the required value of $n_L/s$ along with Eqs. (\ref
{10}, \ref{reheq}), we can express the $T_{r}$ sufficient to generate
the observed matter asymmetry
in terms of the symmetry breaking scale $M$
and the inflaton mass $m_{\rm inf}$:
\begin{equation}\label{trmin}
T_r\gtrsim1.6\times10^{7}{\rm\ GeV}\left(\frac{10^{16}{\rm\
GeV}}{M}\right)^{1/2}\left(\frac{m_{\rm inf}}{10^{11}{\rm\ GeV}}\right)^{3/4}
\left(\frac{0.05\rm{\ eV}}{m_{\nu3}}\right)^{1/2}\,. 
\end{equation}
We show the lower bound on $T_{r}$ calculated using this
equation in Fig. \ref{tvsk} (taking $m_{\nu 3}$ $=0.05$ eV).\footnote{Note that 
the cosmological bound on the sum of the neutrino masses leads to the limit $m_{\nu
3}\lesssim0.2$ eV \cite{Spergel:2006hy}.} The limit in Eq. (\ref{trlimit})
is saturated at $\kappa\approx3\times10^{-7}$, where $m_{\rm inf}=2M_i$. For smaller values of $\kappa$,
sufficient lepton asymmetry cannot be obtained unless the asymmetry is enhanced by
having quasi-degenerate neutrinos.

\begin{figure}[t] 
\includegraphics[angle=0, width=12cm]{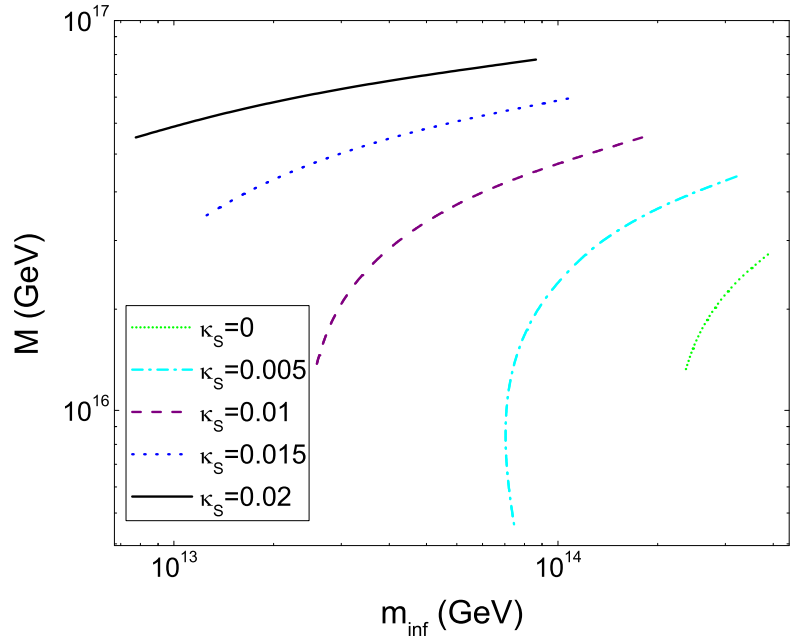} 
 \vspace{-1cm} 
\caption{The inflaton mass $m_{\rm inf}$ vs. the symmetry breaking scale $M$ for smooth hybrid inflation. Only those sections satisfying
$M_*>\sigma_0$ and $0.9<n_s<1.02$ are shown.} \label{minf2}
\end{figure}

\begin{figure}[b] 
\includegraphics[angle=0, width=12cm]{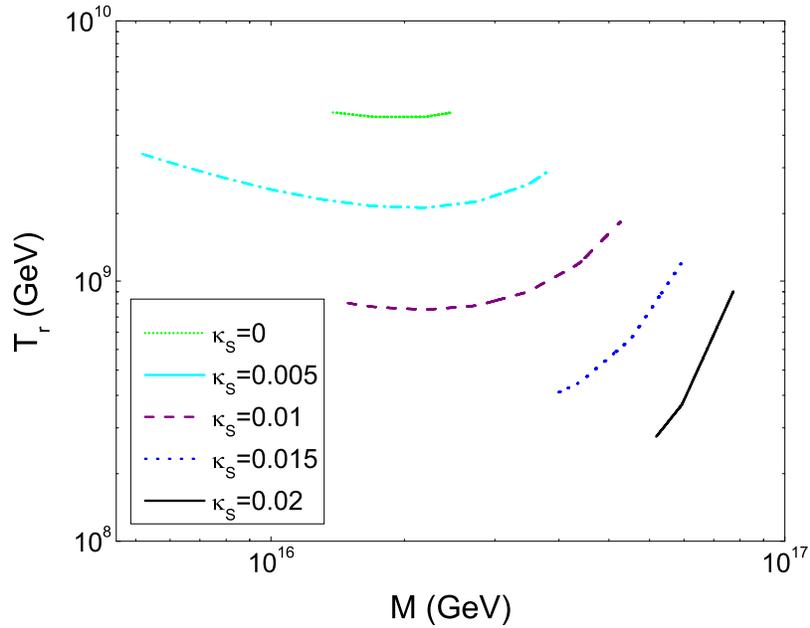}
\vspace{-1cm} 
\caption[The lower bound on the reheat temperature $T_r$ vs. the symmetry breaking scale $M$ for smooth hybrid inflation.]{The lower bound on the reheat temperature $T_r$ vs. the symmetry breaking scale $M$ for smooth hybrid inflation. Only those sections satisfying
$M_*>\sigma_0$ and $0.9<n_s<1.02$ are shown.} \label{sm_mtr}
\end{figure}

For smooth hybrid inflation, $m_{\rm inf}$ is given by 
$2\sqrt{2}v^2/M$.
The value of $m_{\rm inf}$ is shown in Fig. \ref{minf2}.
From Eq. (\ref{reheq}), $M_i/T_r$ is about 10 (40) for $\kappa_S=0$ ($0.01$).
We show the lower bound on $T_r$ (taking $m_{\nu3}=0.05$ eV) 
in Fig. \ref{sm_mtr}.  

\clearpage

\clearpage

An important constraint on supersymmetric inflation models
arises from considering the reheat temperature $T_{r}$
after inflation, taking into account the gravitino problem which requires that
$T_{r}\lesssim10^6$--$10^{10}$ GeV \cite{Khlopov:1984pf}. This
constraint on $T_r$ depends on the SUSY breaking mechanism and the
gravitino mass $m_{3/2}$. For gravity mediated SUSY breaking models
with unstable gravitinos of mass $m_{3/2}\simeq0.1$--1 TeV,
$T_r\lesssim10^6$--$10^9$ GeV \cite{Kawasaki:1995af}, while
$T_r\lesssim10^{10}$ GeV for stable gravitinos \cite{Bolz:2000fu}.
In view of these bounds, smooth hybrid inflation is
relatively disfavoured compared to SUSY hybrid inflation
since $T_r\gtrsim10^9$ for $M=M_{\rm GUT}$.\footnote{A new inflation model related to
smooth hybrid inflation is discussed in \cite{Senoguz:2004ky}
(see also \cite{Asaka:1999jb}), where
the energy scale of inflation $v$ is lower and consequently lower
reheat temperatures are allowed.}

Besides the thermal production of gravitinos which puts an upper bound on $T_r$,
there are also constraints from gravitinos directly produced by inflaton decay.
It was recently pointed out that these constraints can be rather severe for
SUSY and smooth hybrid inflation \cite{Kawasaki:2006gs},
although since the gravitino production depends on the SUSY breaking sector the
models are still viable. As displayed in Fig. \ref{mvsinf}, significantly lower values of $m_{\rm inf}$ can be
obtained with a non-minimal K\"ahler potential for SUSY hybrid inflation. This extends
the allowed range of parameters where the gravitino constraint can be evaded.
For smooth hybrid inflation $m_{\rm inf}$ tends to be higher (Fig. \ref{minf2}).

Finally we note that our estimates for the reheat temperature and matter asymmetry
may be affected due to MSSM flat directions delaying the thermalization of inflaton decay
products or dominating the energy density of the Universe \cite{Allahverdi:2005fq}, although it has been
argued that the flat directions can decay rapidly due to non-perturbative effects \cite{Olive:2006uw}.
Also, there can be additional sources of baryon asymmetry such as `coherent baryogenesis'
\cite{Garbrecht:2003mn}.

\section{Conclusion}\label{conc}
We considered supersymmetric hybrid inflation and 
smooth hybrid inflation models using a general (non-minimal) K\"ahler potential.
The parameter space of the models are extended compared to the minimal
K\"ahler potential case. With a negative mass term in the potential, it is
possible to obtain values of the spectral index in the central WMAP3 range.
Also, sufficient matter asymmetry can be generated with lower values of
the reheat temperature.

In most of the parameter range we consider, the potential develops a false
minimum at large field values and successful inflation is then only possible with
specific initial conditions. However, since these initial conditions lead to eternal inflation,
it is not clear whether this parameter range is
less likely to be observed than the minimal K\"ahler potential case.
Even if we only consider the range for which the potential is monotonic,
it is still possible to obtain a spectral index as low as 0.95 with a negligible tensor to
scalar ratio. For supersymmetric hybrid inflation this requires
$\kappa\gtrsim0.1$ while the gravitino problem favors smaller values of $\kappa$.

\section*{Acknowledgments}

This work is supported in part by the DOE Grant \# DE-FG02-91ER40626 (Q.S.), 
the Bartol Research Institute (M.R. and V.N.{\c S}.),
and  a University of Delaware fellowship
(V.N.{\c S}.). The authors thank George Lazarides, Arunansu Sil
and Mar Bastero-Gil for valuable discussions.

\end{document}